\documentclass[letter]{iopart}
\usepackage{epsfig}

\jl{6}        
\eqnobysec    

\def\beq{\begin{equation}}
\def\eeq{\end{equation}}
\def\rmd{{\rm d}}

\begin{document}

\title[The Speciality Index as invariant indicator  in the BKL Mixmaster Dynamics.
]
{The Speciality Index as invariant indicator  in the BKL Mixmaster Dynamics.}

\author{
Christian Cherubini${}^{\ddag,\,\ast\,\S}$,
Donato Bini${}^{\dag\,\ast, \diamond }$,
Marco Bruni${}^{\S }$,
Zoltan Perjes${}^{\P}$
}
\address{
  ${}^{\ddag}$\
 Faculty of Engineering, University Campus Bio-Medico of Rome,  00155 Rome, Italy.
}
\address{
  ${}^{\ast}$\
International Center for Relativistic Astrophysics, I.C.R.A., 
University of Rome ``La Sapienza,'', I--00185 Rome, Italy.}
\address{
  ${}^{\S}$\
Institute of Cosmology and Gravitation, University of Portsmouth,
Portsmouth, PO1 2EG, United Kingdom.
}
\address{
  ${}^{\dag}$\
Istituto per le Applicazioni del Calcolo \lq\lq M. Picone\rq\rq, C.N.R.,
   I-- 00161 Roma, Italy.
}
\address{
  ${}^{\diamond}$\
Sezione INFN di Firenze, Polo Scientifico, Via Sansone 1,
I--50019 Sesto Fiorentino (FI), Italy.
}
\address{
  ${}^{\P}$\
KFKI Research Institute for Particle and Nuclear Physics, H-1525, Budapest 114, P.O.B. 49, Hungary.
}

\begin{abstract}
The speciality index, 
which has been mainly used 
in Numerical Relativity for studying gravitational waves phenomena as an indicator of the special or non-special Petrov type character of a spacetime, is applied here in the context of Mixmaster cosmology, using the Belinski-Khalatnikov-Lifshitz  map. Possible applications for the associated chaotic dynamics are discussed.

\end{abstract}

\section{Introduction}
The speciality index (SI)~\cite{BC} is an invariant and dimensionless indicator of the special or non-special Petrov algebraic  character of a given spacetime. 
This quantity has been mostly used in Numerical Relativity, for instance to study  a 
black hole which is radiating gravitational waves, or black holes merging~\cite{BC,beetle}.
In this Letter we use  the SI in the context of Cosmology, to study the Belinski-Khalatnikov-Lifshitz (BKL) map~\cite{BK0,BK,BK2}, defining in this way an invariant indicator of the Petrov type \lq\lq fluctuations\rq\rq close to the singularity of the Mixmaster solution.
Possible applications in the context of the associated chaotic dynamics are discussed.

\section{Petrov classification: speciality index.}
The algebraic properties of curvature are a
very useful tool to obtain powerful insights into the character of a given spacetime metric.
In particular those of the Weyl tensor, which is the
trace free part of the Riemann tensor and in vacuum coincides
with it, play  a central role in Einstein's General Relativity theory.
The Penrose-Debever equation $ l_{[u} C_{p]qr[s}l_{t]}l^q l^r=0$
 states the existence of four distinct null eigenvectors for
the most general spacetime: these are known as ``{\it principal
null directions}" (PND)~\cite{kraetal}. 
If all the PND result distinct one has the algebraically general case (Type I).
When some of them coincide, this gives
rise to the algebraically special case summarized as follows:
Type II (one pair of PND coincides),  Type D (two pairs of PND coincide), 
Type III (three PND coincide), Type N(all four PND coincide) and
Type O\,(no PND, because of conformal flatness).

Defining the complex tensor 
$\tilde C_{abcd}=C_{abcd}-i{}^*C_{abcd}$, one can introduce the two
complex curvature invariants
\begin{equation}
I=\frac 1{32}\tilde{C}_{abcd}\tilde{C}^{abcd}\,,\qquad J=\frac 1{384}\tilde{C}_{abcd}\tilde{C}^{cd}{}_{mn}\tilde{C}^{mnab}\,.
\end{equation}
These can be used to define the speciality index~\cite{BC} 
\begin{equation}
\mathcal{S}=27J^2/I^3
\end{equation}
which marks, in an invariant way,
the transition from  certain algebraically special solutions ($\mathcal{S}=1$) 
and the general Petrov type $I$ ($\mathcal{S}\neq 1$)~\cite{kraetal}. 
We point out that for some spacetimes this quantity might be not well defined~\cite{beetle} because 
of the possible vanishing of $I$ and/or $J$, although for the 
vacuum Kasner spacetime~\cite{LL}, as shown in the following, this is not the case. 
This solution of Einstein equations is given by 
\begin{equation}
\rmd s^2=\rmd t^2-t^{2p_1}\rmd x^2-t^{2p_2}\rmd y^2-t^{2p_3}\rmd z^2\,,\quad  \label{LLLL}
\end{equation}
where 
\begin{equation}
p_1+p_2+p_3=p_1^2+p_2^2+p_3^2=1 \label{constr}
\end{equation}
and the indices can take values in the interval $[-\frac{1}{3},1 ]$ only.
The Kasner metric admits two special subcases when two of the $p_i$ indices are equal: it then follows from (\ref{constr}) that either $p_1=p_2=0$, $p_3=1$ (and permutations) and the spacetime is flat in this case, or  $p_1=-1/3$, $p_2=p_3=2/3$ (and permutations) and one has the Kasner locally rotational symmetric type $D$ solution, with a spindle-like
singularity~\cite{kraetal,DIN}.  For other choices of the parameter the Kasner spacetime is of Petrov type I.
A simple calculation, using the constraints listed above, shows~\cite{cbbp1} that the SI for the vacuum Kasner spacetime is
\begin{equation}
\mathcal{S}=-\frac{27}{4}p_1p_2p_3\equiv\frac{27}{4}p_3^2(1-p_3)\,.\label{PIC}
\end{equation}
By direct inspection this quantity ranges from the value $\mathcal{S}=1$ (the type D case) to $\mathcal{S}=0$ (the flat case) and is well defined with continuity for any value of the parameter $p_3$.

\section{Applications to BKL dynamics}

\begin{figure}[tbph]
\begin{center}
\leavevmode
\epsfbox{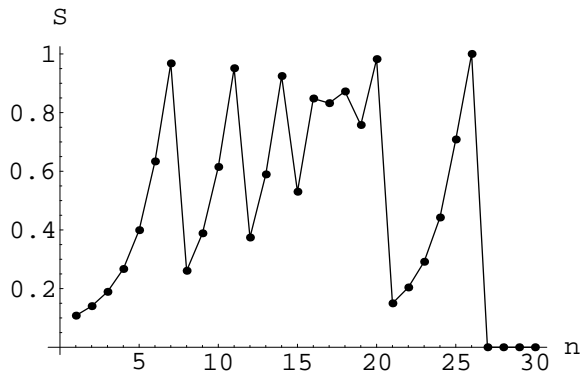}
\end{center}
\caption{{\protect\small Speciality Index for the vacuum BKL map with $%
u_1=7.2328$ in terms of epochs $n$. Interpolation for visual clarification
only. The flattening on the right is due to the map sensibility.} }
\label{FIG1}
\end{figure}

The class of algebraically general Kasner spacetimes contains only algebraically special subcases 
of either type D and type O, and its SI has the simple time independent form (\ref{PIC}).
This  allows us to use the latter to describe the 
Mixmaster dynamics as approximated by BKL Kasner epochs~\cite{BK0,BK,BK2,LL}.
By using the standard parametrization: 
\begin{equation}
\fl \quad p_1=-\frac u{u^2+u+1}\,,\quad p_2=\frac{u+1}{u^2+u+1}\,,\quad p_3=\frac{%
u(u+1)}{u^2+u+1}\,,\qquad u\in [1,+\infty) 
\end{equation}
satisfying the ordering 
\begin{equation}
-\frac 13\le p_1\le 0,\quad 0\le p_2\le \frac 23,\quad \frac 23\le p_3\le
1\,,  \label{ordered}
\end{equation}
 with the sequence of Kasner epochs
given by the rule (Gauss map): 
\begin{eqnarray}
&&u_{n+1}=u_n-1\,\quad \hbox{\rm if}\quad 2\le u_n<\infty \,, \\
&&u_{n+1}=\frac 1{u_n-1}\,\quad \hbox{\rm if}\quad 1\le u_n\le 2\,,
\end{eqnarray}
we obtain from (\ref{PIC}) an ``$n$''-dependent speciality index labelled by each epoch 
\begin{equation}
\mathcal{S}_n=\frac{27}4\,\frac{u_n^2(u_n+1)^2}{(u_n^2+u_n+1)^3}\,.
\end{equation}
Although irrational initial values $u_1$ only should be considered~\cite{LL}, in numerical simulations this requirement results clearly an abstraction, because truncated rational numbers only can be handled.
Using as an example the sequence given by Berger~\cite{Berger} which starts
``close'' to the flat spacetime configuration with $u_1=7.2328$, we get the
speciality index evolution shown in Figure~(\ref{FIG1}). 

We point out that as soon as the system gets close to the type D configuration, it moves rapidly
towards to the type O region, and then gently evolves until it gets close to the type D
case again, and so on.
The flattening of the trend in Figure~(\ref{FIG1}) is due to the well known sensibility of the Gauss map which can generate very long periods of monotonic behavior until oscillations start again.

Pictorially, using the representation of the Mixmaster 
as the motion on a contracting triangular potential well in the time 
direction moving away from the initial singularity, the motion close
to the type D case corresponds to an almost perpendicular bounce on the 
middle of the side of the triangle,  with the incoming free-motion phase before the 
bounce corresponding to the flat space Kasner indices 
and the outgoing free-motion phase after the bounce  corresponding to the Kasner 
indices. 

The bounce is equivalent to a transition from  the one set of Kasner 
indices to the other, for the asymptotic behavior away from the straight 
wall but still far from its time spent at the opposite end in the 
\lq\lq channel\rq\rq corner of the potential where the exact Taub solution 
originates and then returns, and space curvature effects remain 
important since the system point is always close to the potential walls 
of the channel. 

We point out that the BKL parametrization  has cancelled any
information concerning the specific direction in which the motion is
happening, leaving in our case an invariant dynamics in an \lq\lq abstract Petrov space\rq\rq. 

\section{Conclusions}
We have introduced the speciality index in Mixmaster BKL dynamics, in analogy with its use in the numerical treatment of gravitational wave sources.
Because of its gauge invariant nature, time independence  and adimensionality, the Kasner  
 SI and its derived BKL version  can be used in the sophisticated numerical
simulations of Mixmaster, to get useful invariant information concerning chaos~\cite{Barrow,Levin}. 
In particular it would be useful to study the probability associated on the various regions of the segment $[0,1]$ in which the SI ranges during its evolution.  
As pointed out in a recent review on the subject~\cite{Berger2} (see also references therein),
\lq\lq a remaining open question is how closely an actual Mixmaster evolution is approximated by a single BKL sequence.\rq \rq
A direct comparison of the BKL SI (which approximates the Mixmaster dynamics) with the corresponding exact Bianchi IX one will be the appropriate way for approaching the problem numerically, to give an answer to this question.

\section*{Acknowledgments}
The authors thank Robert T. Jantzen and Giovanni Montani for useful suggestions and comments.
CC would like to acknowledge Bruce Bassett and Beverly K. Berger for very stimulating discussions.

\end{document}